\documentclass[12pt]{article}
\usepackage{amssymb,amsmath,epsfig}

\begin{document}

\title{\bf Inflationary Solution of Hamilton Jacobi Equations during Weak Dissipative Regime}
\author{Rabia Saleem\thanks{rabiasaleem@cuilahore.edu.pk} and Muhammad Zubair \thanks{drmzubair@cuilahore.edu.pk}\\
Department of Mathematics, COMSATS University Islamabad, \\Lahore Campus, Pakistan.}

\date{}
\maketitle

\begin{abstract}
In this paper, an elegant mathematical approach is introduced to solve the equations of warm inflationary
model without using extra approximations other than slow-roll. This important inflationary method known
as Hamilton-Jacobian formalism. Here tachyon field and the imperfect fluid are considered to be the cosmic
ingredients to create inflation. A general formalism is developed
for the considered inflationary model and further work is restricted to weak dissipative regime.
 A detailed analysis of the model is presented for
three different choices of bulk and dissipative coefficients taking as constant as well as variable
(function of Hubble parameter and inflaton). In each case, the involved model parameters are constrained
to plot the physical acceptable range of scalar spectral index and tensor to scalar ratio. The parametric
trajectories proved that the acquired results for all the three cases are compatible with Planck astrophysical
data. Furthermore, the existence of warm inflation and slow-roll limit are also verified graphically.
\end{abstract}
{\bf Keywords:} Cosmic Inflation; Cosmological Perturbations; Slow-roll approximation.\\
{\bf PACS:} 98.80.Cq; 05.40.+j.

\section{Introduction}

In reference \cite{1}, Guth put forward a compelling research phenomenon in the field of modern
cosmology named \textit{cosmic inflation}. This theoretical framework became the most successful
for describing the rapid expansion of very early cosmic stage as well as solves some shortcomings
of the hot big-bang model, like ``the horizon problem, the flatness problem and the monopole problem"
\cite{2,3}. The observed anisotropies in the \textit{cosmic microwave background radiation (CMBR)} are
in good agreement with approximately Gaussian, with a scale-invariant primordial power spectrum,
adiabatic scalar perturbation \cite{1a}. Inflationary theory has ability to bring about a causal
mechanism to describe the \textit{large scale structure (LSS)} of the universe and also the source
of the \textit{CMB} anisotropies, since inflaton's quantum fluctuations during the inflationary
expansion are responsible to generate primordial density perturbations \cite{4}.

The inflationary models have two distinct realizations: ``cold (isentropic) inflation"
and ``warm (non-isentropic) inflation" (WI) \cite{5}-\cite{7}. During first type of inflation,
the potential term (of the inflaton's field) remains large as compared to kinetic energy.
Ultimately, this phase terminates with a reheating era that produces radiation into the universe.
Moreover, all interactions of the inflaton field with other fields present in the system are
typically ignored. In contrast to standard ``cold inflation", the other picture of inflation
(\textit{i.e.}, WI) has an essential characteristic of avoiding a reheating period as the
accelerated expansion is ended due to the decay of the inflaton into radiation and relativistic
particles during slow-roll period. The evolution equation of the inflaton field contains dissipative
term originated from this interaction. However, the source of the density fluctuations is the major difference
between these two pictures. During WI scenario, a thermalized radiation component
is present with a restriction $\mathcal{T}_r>H$ (where $\mathcal{T}_r,~H$ be the temperature of
thermal bath and the Hubble expansion rate). Generally, the thermal fluctuations
are produced in spite of quantum \cite{5}-\cite{7}. Bartrum et al. \cite{5a} and Bastero-Gill et al. \cite{5b}
discussed the importance of being warm during inflation and warm little inflation, respectively.

Fluctuation and dissipation phenomena could potentially play an important role in the early universe
cosmology. When matter content of the universe can be split into a subsystem interacting with a
large energy reservoir, then physical processes may be represented through effective dissipation
and stochastic noise terms. Various physical systems have been proposed for the early universe which are
well suited for such a treatment. A treatment involving fluctuation-dissipation dynamics can be implemented
at different levels of coarse graining of the degrees of freedom. In WI, the transfer of inflaton
energy to the radiation bath is mediated by the coupling (dissipation) term in the inflaton's conservation
equation. Due to inner couplings in the radiation fluid itself, an additional effect can arise. Internal
dissipation within radiation fluid slightly disturbs it from thermal equilibrium. Thus, the radiation fluid
behaves as a non-ideal fluid and viscosity effects cannot be neglected. The relevant viscous effect, at the
background level, is due to bulk pressure as it is the only viscous effect appearing in the background equations
\cite{7j}. Decay of massive particles within fluid is an entropy-producing scalar phenomenon, while bulk viscous
pressure $(\Pi)$ has entropy-producing property. The discussion of bulk viscous effects in cosmology,
particularly in inflation, is focused mainly on the effect of $\Pi$ as a negative pressure \cite{9j}. There has
been a surge of interest to study the effects of $\Pi$ which acts as the origin of the accelerating cosmic expansion
\cite{10j}. Tachyon WI with bulk viscous pressure is behaved as an attractor under particular conditions.

As mentioned earlier, inflaton decays during WI and relativistic particles are produced which
usually taken as radiation. By considering the generation of other mass particles in the fluid
could alter the inflationary dynamics by modifying the pressure of fluid in two ways \cite{34}:
firstly, the \textit{hydrodynamic equilibrium pressure} shifts from $P=\frac{\rho}{3}$ to $P=(\gamma-1)\rho$
($1\leq\gamma\leq2$ denotes \textit{adiabatic index}); secondly, taking into account
\textit{non-equilibrium viscous pressure} during inter-particle interaction and particle decay
inside the fluid \cite{35}. The adiabatic index, $\gamma=\frac{4}{3}$, for a quasi-equilibrium high temperature
thermal bath as an inflationary fluid. Misner \cite{M} was probably the first to introduce the
viscosity from the standpoint of particle physics; see also Zel'dovich and Novikov \cite{ZN}. Nevertheless,
on a phenomenological level, the viscosity concept was actually introduced much earlier, with
the first such work being that of Eckart \cite{E}. When considering deviations from thermal equilibrium
to the first order in the cosmic fluid, one should recognize that there are in principle two different
viscosity coefficients, namely the bulk viscosity and the shear viscosity. In view of the commonly
accepted spatial isotropy of the universe, one usually omits the shear viscosity. This is motivated
by the WMAP \cite{WMAP} and Planck observations \cite{P}, and is moreover supported by theoretical
calculations, which show that in a large class of homogeneous and anisotropic universes isotropization
is quickly established. Brevik et al. \cite{B} used a theoretical approach to
provide information concerning quantities related to observations, giving estimations on the inflationary
observables, as well as on the magnitude of the current bulk viscosity itself. From this analysis,
one can see the important implications and the capabilities of the incorporation of viscosity,
which make viscous cosmology a good candidate for the description of Nature.

After introducing WI, several work has been done in this direction. Fang \cite{8} firstly proposed
the concept of coincident particle production during WI and motivated to develop the inflationary
scenario using the condition $\mathcal{T}_r>H$. Moss \cite{9} and further going into detail Yokoyama
and Maeda \cite{10} performed the inflationary calculations including a dissipative term $\Omega\dot{\phi}$
into the evolution equation of the inflaton field. del Campo and Herrera \cite{11} investigated the ``generalized
Chaplygin gas (GCG)" inspired WI driving by an inflaton field containing canonical kinetic term and using
dissipative coefficient, \textit{i.e.}, $\Omega\propto\mathcal{\phi}^m$. The consistency of WI with observational
data is examined using the chaotic potential in the framework of ``loop quantum cosmology"
by Herrera \cite{12}. Herrera \textit{et al}. \cite{13} studied the evolution of generalized dissipative coefficient $\Omega\propto\frac{T^{m}}{\mathcal{\phi}^{m-1}};~m=1,0,-1,3$ during ``intermediate" and ``logamediate" eras. Bamba et al. \cite{16r}
considered single and multiple scalar field theories, tachyon scalar theory and holographic dark energy as models
for current acceleration with the features of quintessence/phantom cosmology, and demonstrated their equivalence
to the corresponding fluid descriptions. Further, WI driven
by a \textit{tachyonic}, \textit{vector} and \textit{non-Abelian gauge} fields is analyzed by Setare and Kamali,
they assumed the scale factor evolves according to ``intermediate" and ``logamediate" models \cite{14}-\cite{16}.
Furthermore, special attention is paid to the equivalence of different dark energy models. Setare and Kamali \cite{12j}
for the first time considered warm tachyon inflation with viscous pressure motivated by the fact that it gives
an end for tachyon inflation.

Sharif and Saleem \cite{17} discussed inflationary dynamics inspired by GCCG (``generalized cosmic Chapygin gas")
using standard and tachyonic fields in ``intermediate" and ``logamediate" scenarios. The same authors presented a
detailed analysis on the dynamics of warm viscous inflation taking isotropic and an anisotropic universe
describing by \textit{Bianchi I} model \cite{18}-\cite{21}. They studied the model for various types of
$\Omega$ (dissipation parameter) and $\xi$ (bulk parameter) and reported that the scalar spectral index $(n_s)$ lies
in the compatible range for less number of e-folds $(N)$. The authors in \cite{22} investigated
the polynomial WI and confirmed the consistency of their results with recent astrophysical data.
Sadjadi and Goodarzi \cite{23} discussed oscillatory type of inflation with \textit{non-minimal kinetic
coupling} as a resolution of few number of e-folds (``non-minimal derivative coupling model \cite{24}").
They reported that the perturbed parameters for this scenario remain compatible with Planck 2013 data.
Extending the previous work, Saleem \cite{25} examined the compatibility of the anisotropic oscillatory
inflation model having \textit{non-minimal kinetic coupling} with Planck 2015 data. However, this type of
inflation does not clear the end stage of inflation that either reheating phase occurs or the universe is
dominated by radiation. In literature, several work has been done on investigating the WI in many alternative
(modified) theories of gravity \cite{26,27}.

However, slow-roll is not the only approach for successful implementation of the cosmic inflationary models,
and particular solutions have been found without using slow-roll limit \cite{28}. Kinney \cite{29} discussed a
general technique in order to evaluate inflationary solutions without implementing the slow-roll
approximation. This technique is mainly based on the notion of considering the scalar field matter's
equation of state as the fundamental part of the dynamical equations, as contrary to the field itself.
This approach is related to the \textit{Hamilton-Jacobi (HJ) formalism} \cite{30}, where the rate of expansion is
considered as the dynamical variable. It is shown that a slow-roll free solution is helpful in calculating the
condition for the model to exit from inflation with inverted-type of potentials, $U(\mathcal{\psi})
=\Lambda^4(1-(\frac{\mathcal{\psi}}{\mu})^{p})$. For early stage of inflation (where $\mathcal{\psi}\ll\mu$),
the slow-roll approximation is taking to be good, but violates well before the ending of inflation \cite{29}.
The same author \cite{29} applied \textit{HJ formalism} to hybrid inflation (more complicated), in this kind
of model, the slow-roll condition fails at all points in the evolution of the inflaton field.

Akhtari \textit{et al.} \cite{29a} considered WI scenario with viscous effects for standard scalar field
using \textit{HJ formalism}. They provided a detail study of the model treating dissipation
and bulk viscous pressure coefficients as constant as well as variable. First case deal with constant
coefficients, which could not portray WI scenario in agreement with Planck observational data for restricted
values of the model parameters. The other two cases for variable coefficients are properly predicted that the
perturbed parameters are in good agreement with Planck data. Motivated by this work, we have applied the
\textit{HJ formalism} on tachyon inspired inflation with viscous pressure. In this scenario, a general criteria is
developed to evaluate the solutions of the tachyonic inflationary model equations given in section \textbf{1}.
Further, the work in this paper is restricted to weak dissipative regime. In section \textbf{2}, the
present model is developed in three different cases, i.e., $(a)~\Omega=\Omega_0,~\xi=\xi_0~
(b)~\Omega=\Omega_0\mathcal{\psi}^m,~\xi=\xi_0~(c)~\Omega=\Omega_0H^2,~\xi=\xi_0\rho$. The involved model
parameters are constrained to plot the $\rho-\mathcal{\psi},~\rho_\mathcal{\psi}-\mathcal{\psi},~
R-n_s,~\mathcal{T}_r-H$ trajectories. The graphical analysis in each case shows that
the tachyon inspired WI with viscous pressure during weak dissipation is in perfect agreement with Planck
data for all constrained model parameters. The results are concluded in the last section.

For simplicity, we took $\hbar=c=\kappa^2=8\pi G=k_{B}=1$, where $G=M^{-2}_{Pl}
,~M_{Pl}=1.2\times10^{19}GeV$ being the Planck mass. The involved model parameters have the units mentioned as
$\mathcal{T}_\gamma\sim H\sim M_{Pl};\quad P,~\rho,~\rho_{\psi}\sim U(\psi)\sim M^{4}_{Pl}$.

\section{General Criteria of Developing an Inflationary Model}

The self-interacting tachyonic scalar field $(\mathcal{\psi})$ and an imperfect fluid
(with bulk viscous pressure) are the components of the assumed matter. The Lagrangian
for tachyon field is given as follows \cite{Macorra}
\begin{equation}\label{1J}
\mathcal{L}=-U(\psi)\sqrt{1-\partial_{\mu}\psi\partial^{\mu}\psi}.
\end{equation}
The considered field has the following energy density and pressure, respectively
\begin{equation}\label{1}
\rho_{\mathcal{\psi}}=\frac{\mathrm{U}(\mathcal{\psi})}{\sqrt{1-\dot{\mathcal{\psi}}^{2}}},\quad
P_{\mathcal{\psi}}=-\mathrm{U}(\mathcal{\psi})\sqrt{1-\dot{\mathcal{\psi}}^{2}},
\end{equation}
where $\mathrm{U}(\mathcal{\psi})$ is the associated effective potential. Important characteristics
of this potential are $\mathrm{U}'(\mathcal{\psi})<0$ and $\mathrm{U}(\mathcal{\psi})\rightarrow0$
as $\mathcal{\psi}\rightarrow\infty$ \cite{31}. The energy density of the imperfect fluid is defined by
$\rho=\mathcal{T}S(\mathcal{\psi},\mathcal{T})$ with temperature $\mathcal{T}$ and entropy density
$S$ \cite{32}; while total pressure of the fluid becomes $P+\Pi$, where the bulk viscous pressure is
expressed as $\Pi=-3\xi H$ \cite{33}.

The dynamical equation for the spatially flat FRW metric is expressed as
\begin{equation}\nonumber
H^2(\mathcal{\psi})=\frac{1}{3}(\rho_{\mathcal{\psi}}+\rho).
\end{equation}
Since the tachyonic inflaton field interacts with the other existed fields and it decays
into the fluid with rate $\Omega$, therefore, the conservation equations can be written
as under
\begin{equation}\label{2}
\dot{\rho_{\mathcal{\psi}}}+3H(\rho_{\mathcal{\psi}}+P_{\mathcal{\psi}})=-\Omega
\dot{\mathcal{\psi}}^2,\quad
\dot{\rho}+3H(\gamma\rho+\Pi)=\Omega\dot{\mathcal{\psi}}^2.
\end{equation}
The coefficient $\Omega$, being the positive quantity, can be dependent upon
temperature and scalar field, i.e., $\Omega\sim \frac{\mathcal{T}^3}{{\mathcal{\psi}}^2}$
\cite{7}. Putting values of $\rho_{\mathcal{\psi}}$ and $P_{\mathcal{\psi}}$, the first
conservation equation becomes
\begin{equation}\label{3}
\frac{\ddot{\mathcal{\psi}}}{1-\dot{\mathcal{\psi}}^2}+3H\dot{\mathcal{\psi}}
+\frac{\mathrm{U}'}{\mathrm{U}}=-\frac{\Omega\mathcal{\psi}}{\mathrm{U}}
\sqrt{1-\dot{\mathcal{\psi}}^2},
\end{equation}
where dot shows time derivative while derivative with respect to $\mathcal{\psi}$ is denoted
by prime. During slow-roll era, the scalar energy density is related to effective potential as
$\rho_{\mathcal{\psi}}\sim \mathrm{U}(\mathcal{\psi})$.
Under slow-roll limits, $\dot{\mathcal{\psi}}\ll1;~\ddot{\mathcal{\psi}}\ll(3H+\frac{\Omega}
{\mathrm{U}})\dot{\mathcal{\psi}}$, the above dynamic equation reduced to the following form
\begin{equation}\label{4}
3H(1+r)\dot{\mathcal{\psi}}=-\frac{\mathrm{U}'}{\mathrm{U}};\quad r=\frac{\Omega}{3H\mathrm{U}}.
\end{equation}
The quasi-stable radiation production restricts the derivative of energy density as $\dot{\rho}\ll
3H(\gamma\rho+\Pi)$ and $\dot{\rho}\ll\Omega\dot{\mathcal{\psi}}^2$, then the energy density
of imperfect fluid could be estimated from second conservation equation as under
\begin{equation}\label{5}
\rho=\gamma^{-1}(Q\dot{\mathcal{\psi}}^2-\Pi); \quad Q=\frac{\Omega}{3H}.
\end{equation}

In canonical WI scenario, the strength of $\Omega$ (the thermal damping) should be relatively
compared to $H$ (Hubble expansion damping). We must analyze the WI model in background and
linear perturbation levels on our expanding over timescales, which are shorter than the variation of
expansion rate, but large compared to the microphysical processes
\begin{equation}\label{6}
\frac{\mathrm{U}}{\Omega}\ll\tau\ll H^{-1}\quad\Rightarrow\quad \Omega\gg H\mathrm{U}.
\end{equation}
Putting values of $\rho_{\mathcal{\phi}}$ and $\rho$ in the Friedmann equation, we get
\begin{equation}\label{7}
\dot{H}=-\frac{1}{2}\left(\mathrm{U}(\mathcal{\psi})+\frac{\Omega}{3H}\right)\dot{\mathcal{\psi}}^2
=-\frac{1}{2}\mathrm{U}(\mathcal{\psi})(1+r)\dot{\mathcal{\psi}}^2.
\end{equation}
From the above equation, the term $\dot{\mathcal{\psi}}$ is obtained as follows
\begin{equation}\label{8}
\dot{\psi}=-\frac{2H'(\psi)}{U(\psi)(1+r)}.
\end{equation}
Applying the condition, $\rho_{\mathcal{\psi}}\gg\rho$ on Friedmann equation, we get following expression
of $\mathrm{U}(\mathcal{\psi})$ as
\begin{equation}\label{9}
\mathrm{U}(\mathcal{\psi})=3H^{2}(\mathcal{\psi})\left[1-\frac{2H^2(\mathcal{\psi})
H'^{2}(\mathcal{\psi})}{\mathrm{U}^2(\mathcal{\psi})(1+r)^2}\right],
\end{equation}
which leads to a polynomial of order three in $\mathrm{U}(\mathcal{\psi})$ as under
\begin{eqnarray}\nonumber
9H^2(\mathcal{\psi})\mathrm{U}^3(\mathcal{\psi})&+&(6H(\mathcal{\psi})\Omega(\mathcal{\psi})
-27H^4(\mathcal{\psi}))\mathrm{U}^2(\mathcal{\psi})+(\Omega^{2}(\mathcal{\psi})-18H^3(\mathcal{\psi})
\Omega(\mathcal{\psi}))\\\label{10}&\times&\mathrm{U}(\mathcal{\psi})-3H^2(\mathcal{\psi})\Omega^2
(\mathcal{\psi})+54H^6(\mathcal{\psi})H'^{2}(\mathcal{\psi})=0.
\end{eqnarray}

The most important physically observed parameter is $\epsilon$, mathematically
expressed as
\begin{equation}\label{11}
\epsilon=-\frac{\dot{H}}{H^{2}}=\frac{2}{\mathrm{U}(\mathcal{\psi})(1+r)}\left(\frac{H'(\mathcal{\psi})}
{H(\mathcal{\psi})}\right)^2.
\end{equation}
The fluid energy density can be evaluated using above expression and the expression of $\dot{\mathcal{\phi}}^2$
as
\begin{equation}\label{12}
\rho=\gamma^{-1}\left(\frac{2Q^2}{3\mathrm{U}^2(\mathcal{\psi})(1+r)^2}\rho_{\mathcal{\psi}}\epsilon-\Pi\right).
\end{equation}
Therefore, at the end of inflation and for the case $r\gg1$, the above relation reduced to
\begin{equation}\label{13}
\epsilon=1\quad\Rightarrow\quad\rho=\gamma^{-1}
\left(\frac{2}{3}\rho_{\mathcal{\psi}}-\Pi\right).
\end{equation}
The parameter $N$ can be calculated as
\begin{equation}\label{14}
N(\mathcal{\psi})=-\int^{\mathcal{\mathcal{\psi}}_{e}}_{\mathcal{\psi}}\frac{1}{2}\mathrm{U}(\mathcal{\psi})
(1+r)\frac{H(\mathcal{\psi})}{H'(\mathcal{\psi})}d\mathcal{\psi},
\end{equation}
where $\mathcal{\psi}_{e}$ and $\mathcal{\psi}$ be the start and end value of inflaton. The other
slow-roll parameter is given as under
\begin{equation}\label{15}
\eta=-\frac{\ddot{H}}{H\dot{H}}=\frac{2}{\mathrm{U}(\mathcal{\psi})(1+r)}\frac{H''(\mathcal{\psi})}{H(\mathcal{\psi})}.
\end{equation}

The thermal power spectrum of scalar perturbation is read as \cite{31}
\begin{equation}\label{16}
P_{s}=32\mathcal{T}_{r}\left(\frac{\Omega H}{\mathrm{U}(\mathcal{\psi})}\right)
^{\frac{1}{2}}\frac{\exp[-2\chi(\mathcal{\psi})]}{\mathrm{U}'(\mathcal{\psi})},
\end{equation}
where the quantity $\chi(\mathcal{\psi})$ (auxiliary function) is calculated as \cite{31}
\begin{eqnarray}\nonumber
\chi(\mathcal{\psi})&=&-\int\frac{\frac{\Omega^{\prime}}{\mathrm{U}^{\prime}}}{3H+\frac{\Omega}
{\mathrm{U}}}+\frac{9}{8G(\mathcal{\psi})}\frac{2H+\frac{\Omega}{\mathrm{U}}}
{(3H+\frac{\Omega}{\mathrm{U}})^2}\left[\Omega+4H\mathrm{U}-\left(\frac{\frac{\Omega^{\prime}
\mathrm{U}^{\prime}}{\mathrm{U}}}{12H\gamma(3H+\frac{\Omega}{\mathrm{U}})}\right.\right.
\\\label{17}&\times&\left.\left.\left[(\gamma-1)
+\Pi\frac{\xi_{,\rho}}{\xi}\right]\right)\right]\frac{\mathrm{U}'}{\mathrm{U}^2}d\phi,
\end{eqnarray}
here
\begin{equation}\nonumber
G(\mathcal{\psi})=1-\frac{1}{8H^2}\left(2\gamma\rho+3\Pi+\left(\frac{\gamma\rho+\Pi}{\gamma}\right)
\left[\frac{\xi_{,\rho}}{\xi}\Pi-1\right]\right).
\end{equation}
The power spectrum of tensor perturbation is \cite{31}
\begin{equation}\nonumber
P_{T}=\frac{H^{2}}{2\pi^2}\coth[\frac{k}{2\mathcal{T}}].
\end{equation}
The scalar spectral index $(n_s)$ can be calculated in the following form
\begin{equation}\label{21a}
n_{s}-1=\frac{d\ln P_{s}}{d\ln k}=\left[\frac{\Omega'}{2\Omega}+
\frac{H'}{2H}-\frac{\mathrm{U}'}{2\mathrm{U}}-2\frac{\mathrm{U}''}{\mathrm{U}'}
-2\chi'(\mathcal{\psi})\right]\dot{\mathcal{\psi}},
\end{equation}
where $d\ln k=Hdt$. The tensor-scalar ratio is defined by
\begin{equation}\label{21}
R(k_{0})=\frac{\mathcal{T}_r}{64\pi^2}\left(\frac{\mathrm{U}}{\Omega}\right)^{\frac{1}{2}}H^{\frac{3}{2}}
\mathrm{U}'^2(\mathcal{\psi})\exp[2\chi(\mathcal{\psi})]\coth[\frac{k}{2\mathcal{T}}]\mid_{k=k_{0}}.
\end{equation}

Now, we will analyze the behavior of perturbed model quantities by comparing with
recent Planck data in the following section. To complete the task, a specific power-law
form of Hubble parameter as a function of inflaton is proposed as $H(\psi)=H_0\psi^{n}$,
where $n$ is an arbitrary constant and $H_0$ has dimension $M^{1-n}$.

\section{Weak Dissipative Regime}

Here, we will develop all these calculation under weak dissipation condition,
$r\ll1\Rightarrow \Omega\ll3H$. The coefficients $\Omega$ and $\xi$ are considered
to be constant and variable in alternative cases.

\subsection{{Case I:\quad $\Omega=\Omega_0;\quad \xi=\xi_0$}}

During weak dissipative regime and for constant coefficients, Eq.(\ref{9}) is reduced to be
\begin{equation}\label{50}
\mathrm{U}(\mathcal{\psi})=3H^{2}(\mathcal{\psi})\left[1-\frac{2H'^{2}
(\mathcal{\psi})}{\mathrm{U}^2(\mathcal{\psi})}\right],
\end{equation}
which leads to following polynomial of order three in $\mathrm{U}(\psi)$
\begin{equation}\nonumber
\mathrm{U}^3(\mathcal{\psi})-3H^2(\mathcal{\psi})\mathrm{U}^2(\mathcal{\psi})+6H^2(\mathcal{\psi})H'^{2}(\mathcal{\psi})=0.
\end{equation}
Taking $H=H_0\mathcal{\psi}^n$, the solution of $\mathrm{U}(\mathcal{\psi})$ has the following form
\begin{eqnarray}\nonumber
\mathrm{U}(\mathcal{\psi})&=&H^{7}_0n^6\mathcal{\psi}^{2(2n-1)}[\sqrt{9H^{2}_0\mathcal{\psi}^{8(n+1)}
-6n^2\mathcal{\psi}^{2(5n+2)}}+H^{6}_0n^6\mathcal{\psi}^{6n}-3H^{4}_0n^2\mathcal{\psi}^{4(n+1)}]^{-\frac{1}{3}}
\\\label{51}&+&H^{2}_0n^2\mathcal{\psi}^{2(n-1)}+\mathcal{\psi}^{-2}.
\end{eqnarray}
The value of $\dot{\mathcal{\psi}}^2$ can be calculated from the expression
$\dot{\mathcal{\psi}}^2=-2\dot{H}\mathrm{U}(\mathcal{\psi})$ (Eq.(\ref{8})) as
\begin{eqnarray}\nonumber
\dot{\mathcal{\psi}}&=&-2H'(\mathcal{\psi})[H^{2}_0n^2\mathcal{\psi}^{2(n-1)}+\mathcal{\psi}^{-2}
+H^{7}_0n^6\mathcal{\psi}^{2(2n-1)}[\sqrt{9H^{2}_0\mathcal{\psi}^{8(n+1)}
-6n^2\mathcal{\psi}^{2(5n+2)}}\\\label{52}&+&H^{6}_0n^6\mathcal{\psi}^{6n}-3H^{4}_0n^2
\mathcal{\psi}^{4(n+1)}]^{-\frac{1}{3}}]^{-1}.
\end{eqnarray}

The weak dissipation regime produced the following number of e-folds given in Eq.(\ref{14}) as
\begin{equation}\label{53}
N=-\frac{1}{2}\int^{\mathcal{\psi}}_{\mathcal{\psi}_e}\frac{H(\mathcal{\psi})}{H'
(\mathcal{\psi})}\mathrm{U}(\mathcal{\psi})d\mathcal{\psi}=-\frac{1}{2n}\int^{\mathcal{\psi}}_{\mathcal{\psi}_e}
\mathcal{\psi} \mathrm{U}(\mathcal{\psi})d\mathcal{\psi},
\end{equation}
which leads to the solution of $\mathcal{\psi}$ as
\begin{equation}\label{54}
\mathcal{\psi}=\exp[-[nN+\frac{1}{2}[\ln\mathcal{\psi}_e+\frac{H^{2}_0n}{2}(1+H^3_0n^2)
\mathcal{\psi}^{2n}_e]]+\frac{H^{2}_0n}{4}(1+H^{3}_0n^2)].
\end{equation}
The slow-roll parameters are reduced to
\begin{eqnarray}\nonumber
\epsilon&=&2n^2[1-2H^{3}_0n^2
\exp[-4nN-2[\ln\mathcal{\psi}_e+\frac{H^{2}_0n}{2}(1+H^3_0n^2)\mathcal{\psi}^{2n}_e]\\\nonumber&-&H^{2}_0n(1+H^{3}_0n^2)]],\\\nonumber
\eta&=&[1-n^2H^{2}_0n^2(1+H^{3}_0n^2)
\exp[-[nN-\frac{1}{2}[\ln\mathcal{\psi}_e+\frac{H^{2}_0n}{2}(1+H^3_0n^2)\mathcal{\psi}^{2n}_e]
\\\label{55}&+&\frac{H^{2}_0n}{4}(1+H^{3}_0n^2)]]]3n(n-1).
\end{eqnarray}
The value of $\mathcal{\psi}_e$ can be evaluated using the exit condition $\epsilon=1$ in the following form
\begin{equation}\label{54a}
\mathcal{\psi}_e=[\exp[-\frac{H^{2}_0n}{2}(1+H^3_0n^2)][-2nN+\ln[\frac{2n^2-1}{4H^3_0n^4}]^{-\frac{1}{2}}
-\frac{H^{2}_0n^2}{2}(1+H^3_0n^2)]]^{\frac{1}{1+2n}}.
\end{equation}
\begin{figure}\centering
\epsfig{file=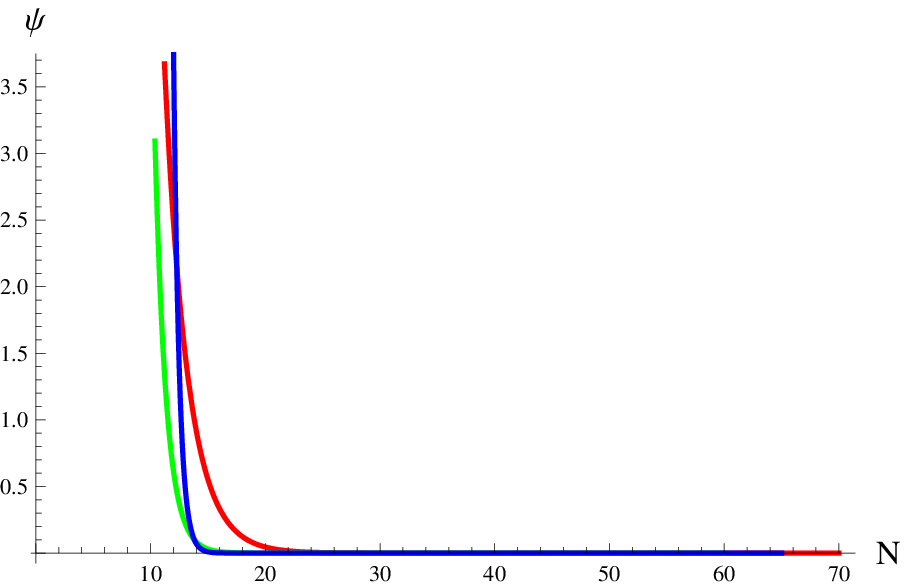,
width=0.55\linewidth}\epsfig{file=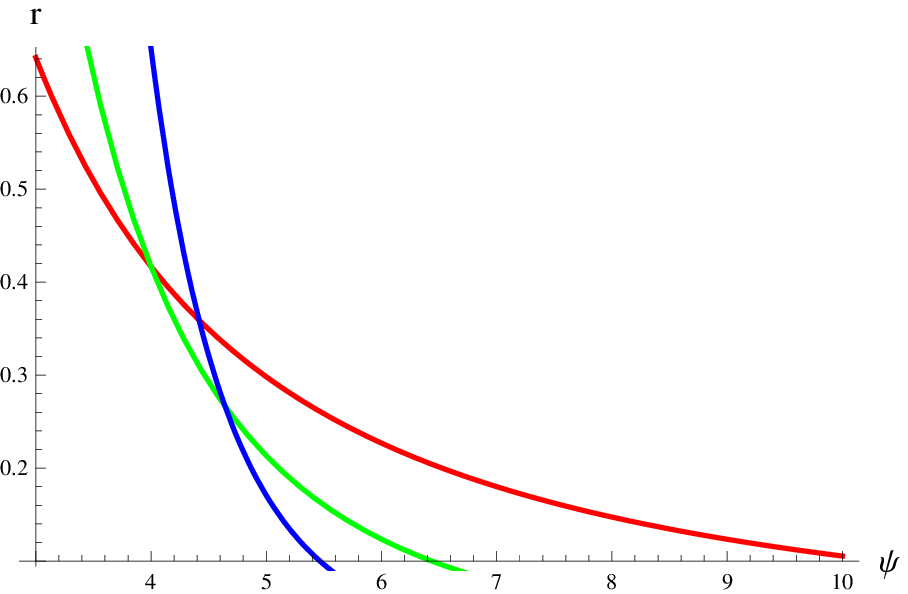,
width=0.55\linewidth}\caption{Left plot $\psi$ versus $N$: Red for $n=0.5,~H_{0}=1.6\times10^{-2}$;
Green for $n=1,~H_{0}=5\times10^{-3}$; Blue for $n=2,~H_{0}=5\times10^{-4}$. Right plot $r$ verses
$\psi$ for Red for $n=0.5,~H_{0}=1.6\times10^{-2}$;
Green for $n=1,~H_{0}=5\times10^{-3}$; Blue for $n=2,~H_{0}=5\times10^{-4}$.}
\end{figure}
Figure \textbf{1} (left plot) shows that tachyon field slowly rolls down to its minimum point
and then attains stable configuration. The right plot of Fig.\textbf{1} satisfies the restriction of
work done in weak dissipative regime as $r\ll1$ for above mentioned values of the model parameters.
The energy density is restricted under $r\ll1$ as
\begin{equation}\label{56}
\rho=\gamma^{-1}\left(\frac{2}{3}\frac{H^{2}(\mathcal{\psi})}{\mathrm{U}^2(\mathcal{\psi})}-\Pi\right)=
\gamma^{-1}\left(\frac{2H^{2}_0\mathcal{\psi}^{2n}}{3H^{6}_0\mathcal{\psi}^{6-4n}+3H^{4}_0n^4
\mathcal{\psi}^{4n-4}+6H^{5}_0n^2\mathcal{\psi}}-\Pi\right).
\end{equation}
\begin{figure}\centering
\epsfig{file=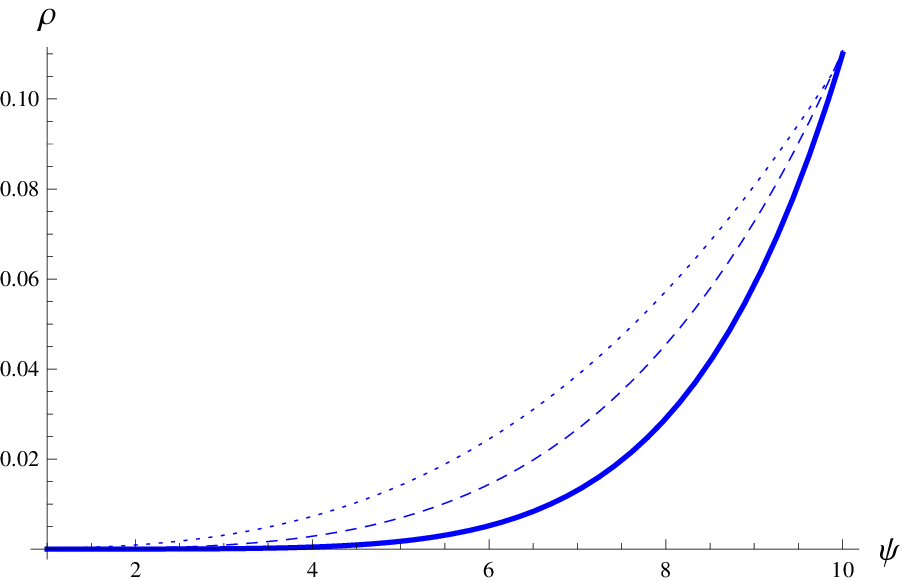,
width=0.55\linewidth}\epsfig{file=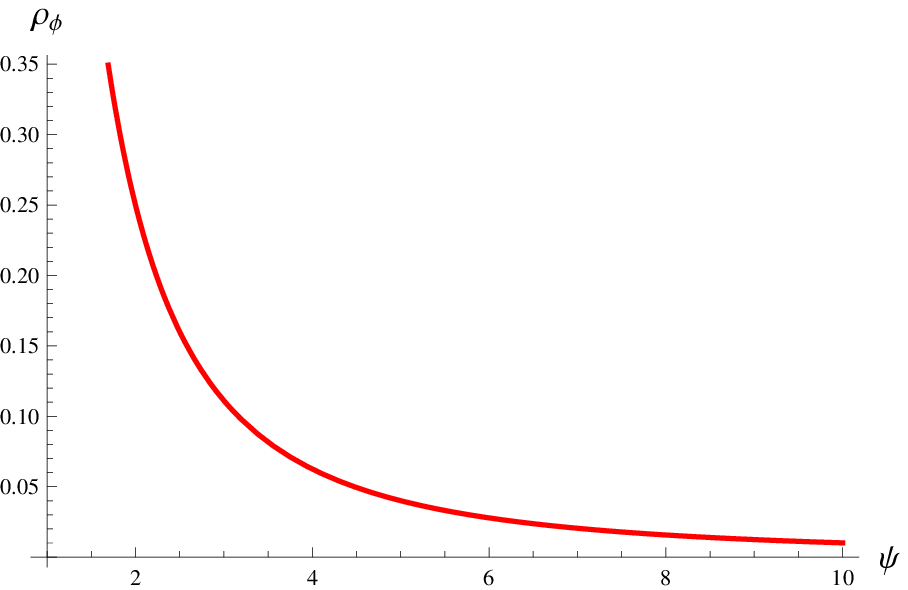,
width=0.55\linewidth}\caption{Left plot $\rho$ versus $\mathcal{\phi}$: Dotted for $n=0.5,~H_{0}=1.6\times10^{-2}$;
Dashed for $n=1,~H_{0}=5\times10^{-3}$; Thick for $n=2,~H_{0}=5\times10^{-4}$. Right plot $\rho_\mathcal{\phi}$
versus $\mathcal{\phi}$: Dotted for $n=0.5,~H_{0}=1.6\times10^{-2}$; Dashed for $n=1,~H_{0}=5\times10^{-3}$;
Thick for $n=2,~H_{0}=5\times10^{-4}$.}
\end{figure}
The plots of $\rho$ and $\rho_\psi$ versus $\psi$ are plotted in left and right plots of fig.\textbf{2}. By comparing
the attained range of both energy densities for specified values of the model parameters, it can be noticed that
the slow-roll condition is true in this scenario. The other involved parameters are fixed to $\gamma=1.5,
~\xi_{0}=7\times10^{-14}M^{3}_{p}.$

The auxiliary function is reduced to
\begin{eqnarray}\nonumber
G(\mathcal{\psi})&=&1-\frac{1}{12}\left[\frac{1}{H^{6}_0\mathcal{\psi}^{6-4n}+H^{4}_0n^4
\mathcal{\psi}^{4n-4}+2H^{5}_0n^2\mathcal{\psi}}
\left(2-\frac{1}{\gamma}\right)\right],\\\label{57}
\overline{\chi}(\mathcal{\psi})&=&-2\ln \mathrm{U}.
\end{eqnarray}
The scalar power spectrum in weak limit has the form as under
\begin{equation}\label{58}
P_{s}=32\mathcal{T}_r\Omega^{\frac{1}{2}}_0H^{\frac{1}{2}}_0\mathcal{\psi}^{\frac{n}{2}}\left[\frac{(\mathcal{\psi}^{-2}+H^{2}_0n^2
(1+H^{3}_0n^2)\mathcal{\psi}^{2(n-1)}+H^{3}_0\mathcal{\psi}^2)^\frac{7}{2}}{(-2\mathcal{\psi}^{-3}+(2n-2)H^{2}_0n^2
(1+H^{3}_0n^2)\mathcal{\psi}^{2n-3}+2H^{3}_0\mathcal{\psi})^2}\right],
\end{equation}
where $\psi$ is given in Eq.(\ref{54}). The tensor power spectrum is calculated to be
\begin{equation}\label{59}
P_{T}=\frac{H^{2}_0}{2\pi^2}\mathcal{\psi}^{2n}\coth[\frac{k}{2\mathcal{T}}].
\end{equation}
The parameter $n_s$ becomes
\begin{eqnarray}\nonumber
n_{s}-1&=&32\Omega^{\frac{1}{2}}_0\mathcal{T}_r H^{\frac{1}{2}}_0\exp[-n][\frac{n}{2}
\mathcal{\psi}^{-1}+\frac{7}{2}(\mathcal{\psi}^{-2}+H^{2}_0n^2(1+H^{3}_0n^2)
\psi^{n-2}+H^{3}_0\mathcal{\psi}^2)^{-1}\\\nonumber&\times&(-2\mathcal{\psi}^{-3}+(n-2)H^{2}_0n^2
(1+H^{3}_0n^2)\mathcal{\psi}^{n-3}+2H^{3}_0\mathcal{\psi})-2(-2\mathcal{\psi}^{-3}+(2n-2)\\\nonumber&\times&H^{2}_0n^2
(1+H^{3}_0n^2)\mathcal{\psi}^{2n-3}+2H^{3}_0\mathcal{\psi})^{-1}(6\mathcal{\psi}^{-4}+(2n-2)(2n-3)H^{2}_0n^2
\\\label{61}&\times&(1+H^{3}_0n^2)\mathcal{\psi}^{2n-4}+2H^{3}_0)].
\end{eqnarray}
The above two equations of spectrum lead to express tensor-scalar spectrum ratio as
\begin{equation}\label{60}
R=\frac{H^{\frac{3}{2}}_0\Omega^{-\frac{1}{2}}_0}{64\pi^2\mathcal{T}_r}\mathcal{\psi}^{\frac{3n}{2}}
\left[\frac{(-2\mathcal{\psi}^{-3}+(2n-2)H^{2}_0n^2(1+H^{3}_0n^2)\mathcal{\psi}^{2n-3}+2H^{3}_0\mathcal{\psi})^2}
{(\mathcal{\psi}^{-2}+H^{2}_0n^2(1+H^{3}_0n^2)\mathcal{\psi}^{2(n-1)}+H^{3}_0\mathcal{\psi}^2)^\frac{7}{2}}\right]
\coth[\frac{k}{2\mathcal{T}}].
\end{equation}
\begin{figure}\center
\epsfig{file=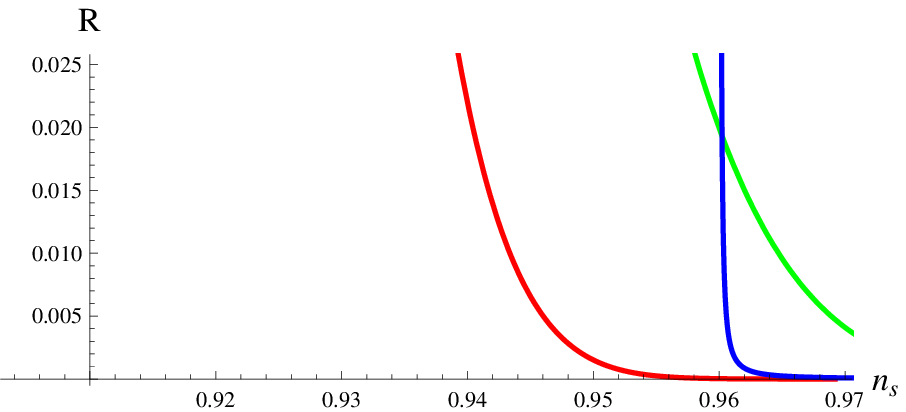,
width=0.55\linewidth}\caption{$R$ versus $n_{s}$: Red for $n=0.5,~H_{0}=1.6\times10^{-2},~\Omega=0.03$;
Green for $n=1,~H_{0}=5\times10^{-3},~\Omega=0.01$; Blue for $n=2,~H_{0}=3\times10^{-3},~\Omega=0.033$.}
\end{figure}
The parametric trajectory of $R-n_s$ is plotted in Fig.\textbf{3} for specified values of model parameters.
These trajectories fall in the physical acceptable range allowed by Planck astrophysical data as $R<0.11$ for
$n_s=0.968$. Hence, the constant coefficients case is compatible with Planck data for constrained values of the
model parameters.

The temperature of weak regime can be expressed as under using the relation, $\rho=\mathcal{T}S=C_\gamma \mathcal{T}^4$
\begin{equation}\label{62}
\mathcal{T}_\gamma=\left[\frac{1}{\gamma C_\gamma}(\frac{2H^{2}_0\mathcal{\psi}^{2n}}{3H^{6}_0\mathcal{\psi}^{6-4n}
+3H^{4}_0n^4\mathcal{\psi}^{4n-4}+6H^{5}_0n^2\mathcal{\psi}}+3\xi_0H_0\mathcal{\psi}^n)\right]^{\frac{1}{4}}.
\end{equation}
\begin{figure}\center
\epsfig{file=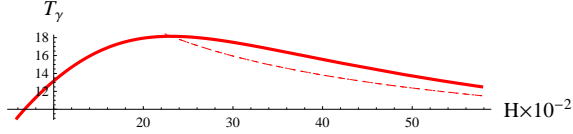,
width=0.55\linewidth}\caption{Parametric plot of $\mathcal{T}_\gamma$ versus $H$: Thick for $n=0.5,~H_{0}=1.6\times10^{-2}$;
Dotted for $n=1,~H_{0}=5\times10^{-3}$; Dashed for $n=2,~H_{0}=6\times10^{-4}$.}
\end{figure}
Figure \textbf{4} verifies that the current model gracefully describes the existence of WI by satisfying the condition
$\mathcal{T}_\gamma>H$ for constrained model parameters. For constant coefficients case, the expression $\Omega<3H$
constrained as follows
\begin{equation}\label{63}
\Omega_0<3H_0[\mathcal{\psi}^{n-2}+H^{2}_0n^2(1+H^{3}_0n^2)\mathcal{\psi}^{3n-2}+H^{3}_0
\mathcal{\psi}^{n+2}].
\end{equation}
Here, $\mathcal{T}_r>H$ and above mentioned conditions come to following inequality,
which holds during inflation as
\begin{eqnarray}\nonumber
3\gamma C_\gamma(\frac{1}{2}H^{8}_0\mathcal{\psi}^{6-2n}&+&H^{7}_0n^2\mathcal{\psi}^{2n+1}
+\frac{1}{2}n^4H^{6}_0\mathcal{\psi}^{6n-4})-9\xi_0
(\frac{1}{2}H^{5}_0\mathcal{\psi}^{6-5n}+H^{4}_0n^2\mathcal{\psi}^{1-n}
\\\label{64}&+&\frac{1}{2}n^4H^{3}_0\mathcal{\psi}^{3n-4})<1.
\end{eqnarray}
The expressions for amplitude of tensor perturbations without and with viscous pressure shall satisfy the following constraint
for $H_0$, respectively
\begin{eqnarray}\nonumber
H_0<\sqrt{2\pi^2r^{\ast}P^{\ast}_s}\exp[-[n^2N+\frac{n}{2}[\ln\mathcal{\psi}_e+\frac{H^{2}_0n}{2}(1+H^3_0n^2)\mathcal{\psi}^{2n}_e]]
+\frac{H^{2}_0n}{4}(1+H^{3}_0n^2)],\\\nonumber H_0<\sqrt{\frac{2\pi^2r^{\ast}
P^{\ast}_s}{\coth[\frac{k}{2\mathcal{T}}]}}\exp[-[n^2N+\frac{n}{2}[\ln\mathcal{\psi}_e+\frac{H^{2}_0n}{2}(1+H^3_0n^2)
\mathcal{\psi}^{2n}_e]]+\frac{H^{2}_0n}{4}(1+H^{3}_0n^2)],\\\label{65}
\end{eqnarray}
where $P_s$ is given in Eq.(\ref{58}).

Next, we will use the same formalism taking variable dissipation coefficient.

\subsection{Case II:\quad $\Omega=\Omega_0\mathcal{\psi}^m;\quad \xi=\xi_0$}

In this case, the expressions for $\psi,~\epsilon,~\eta$ and $\rho$ remains the same as in the previous case.
While for variable dissipation coefficient (as a function of $\mathcal{\psi}$), $\overline{\chi}(\mathcal{\psi})$
is turn out to be
\begin{eqnarray}\nonumber
\overline{\chi}(\mathcal{\psi})&=&\frac{\Omega_0m\mathcal{\psi}^{m-3n+3}}{3(2n-2)(m-3n+3)H^{3}_0n^2
(1+H^{3}_0n^2)}+\ln[\mathcal{\psi}^{-2}+H^{2}_0n^2
(1+H^{3}_0n^2)\\\label{66}&\times&\mathcal{\psi}^{2n-2}+H^{3}_0\mathcal{\psi}^2]+\frac{(\gamma-1)(2n-2)
^2\Omega_0m}{144(m-5n)\gamma H^{5}_0n^2
(1+H^{3}_0n^2)}\mathcal{\psi}^{m-5n}.
\end{eqnarray}
Using above equation, $P_s$ has the form as mentioned below
\begin{eqnarray}\nonumber
P_{s}&=&32T_r\Omega^{\frac{1}{2}}_0H^{\frac{1}{2}}_0\mathcal{\psi}^{\frac{m+n}{2}}[\frac{(\mathcal{\psi}^{-2}+H^{2}_0n^2
(1+H^{3}_0n^2)\mathcal{\psi}^{2(n-1)}+H^{3}_0\mathcal{\psi}^2)^-\frac{1}{2}}{(-2\mathcal{\psi}^{-3}+(2n-2)H^{2}_0n^2
(1+H^{3}_0n^2)\mathcal{\psi}^{2n-3}+2H^{3}_0\mathcal{\psi})^2}\\\nonumber&+&\exp[\frac{-2\Omega_0m\mathcal{\psi}^{m-3n+3}}
{3(2n-2)(m-3n+3)H^{3}_0n^2
(1+H^{3}_0n^2)}-2\ln[\mathcal{\psi}^{-2}+H^{2}_0n^2(1+H^{3}_0n^2)\\\label{67}&\times&\mathcal{\psi}^{2n-2}+H^{3}_0\mathcal{\psi}^2]
-\frac{2(\gamma-1)(2n-2)^2\Omega_0m}{144(m-5n)\gamma H^{5}_0n^2(1+H^{3}_0n^2)}\mathcal{\psi}^{m-5n}]],
\end{eqnarray}
where $\mathcal{\psi}$ is perviously used given in Eq.(\ref{54}). The logarithm derivative of the above equation
leads to following parameter
\begin{eqnarray}\nonumber
n_{s}-1&=&[(\frac{m+n}{2})\mathcal{\psi}^{-1}-\frac{1}{2}[\mathcal{\psi}^{-2}+H^{2}_0n^2(1+H^{3}_0n^2)
\mathcal{\psi}^{2n-2}+H^{3}_0\mathcal{\psi}^2]^{-1}[-2\mathcal{\psi}^{-3}\\\nonumber&+&(2n-2)H^{2}_0n^2(1+H^{3}_0n^2)
\mathcal{\psi}^{2n-3}+2H^{3}_0\mathcal{\psi}^2]-2(-2\mathcal{\psi}^{-3}+(2n-2)H^{2}_0n^2\\\nonumber&\times&(1+H^{3}_0n^2)
\mathcal{\psi}^{2n-3}+2H^{3}_0\mathcal{\psi}^2)^{-1}(6\mathcal{\psi}^{-4}+(2n-2)(2n-3)H^{2}_0n^2(1+H^{3}_0n^2)
\\\nonumber&\times&\mathcal{\psi}^{2n-3}+2H^{3}_0)-\frac{2\Omega_0m\mathcal{\psi}^{m-3n+3}}{3(2n-2)(m-3n+3)H^{3}_0n^2
(1+H^{3}_0n^2)}\\\nonumber&-&2(\frac{-2\mathcal{\psi}^{-3}+(2n-2)H^{2}_0n^2(1+H^{3}_0n^2)\mathcal{\psi}^{2n-3}+2H^{3}_0\mathcal{\phi}}
{\mathcal{\psi}^{-2}+H^{2}_0n^2(1+H^{3}_0n^2)\psi^{2n-2}+H^{3}_0\mathcal{\psi}^2})
\\\nonumber&-&\frac{2(\gamma-1)(2n-2)^2\Omega_0m}{144\gamma H^{5}_0n^2(1+H^{3}_0n^2)}\mathcal{\psi}^{m-5n-1}].
\end{eqnarray}
\begin{figure}\center
\epsfig{file=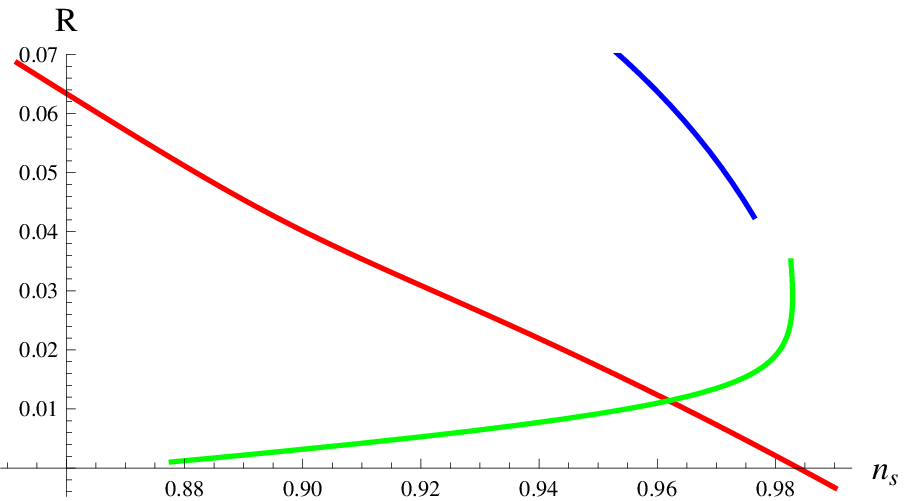,
width=0.55\linewidth}\caption{$R$ versus $n_{s}$: Red for $n=0.5,~H_{0}=3\times10^{-3},~\Omega=0.0015$;
Green for $n=1.5,~H_{0}=4\times10^{-4},~\Omega=0.0012$; Blue for $n=2.5,~H_{0}=3.5\times10^{-2},~\Omega=0.003$.}
\end{figure}
The tensor power spectrum is calculated in Eq.(\ref{59}). Equations (\ref{59}) and (\ref{67}) combined to
produce physical parameter $R$ as
\begin{eqnarray}\nonumber
R&=&\frac{H^{\frac{3}{2}}_0\Omega^{-\frac{1}{2}}_0}{64\pi^2\mathcal{T}_r}\mathcal{\psi}^{\frac{3n-m}{2}}
[\frac{(-2\mathcal{\psi}^{-3}+(2n-2)H^{2}_0n^2
(1+H^{3}_0n^2)\mathcal{\psi}^{2n-3}+2H^{3}_0\mathcal{\psi})^2}{(\psi^{-2}+H^{2}_0n^2
(1+H^{3}_0n^2)\mathcal{\psi}^{2(n-1)}+H^{3}_0\mathcal{\psi}^2)^{-\frac{1}{2}}}]\\\nonumber&\times&
\exp[\frac{2\Omega_0m\mathcal{\psi}^{m-3n+3}}{3(2n-2)(m-3n+3)H^{3}_0n^2
(1+H^{3}_0n^2)}+2\ln[\mathcal{\psi}^{-2}+H^{2}_0n^2(1+H^{3}_0n^2)\\\label{68}&\times&\mathcal{\psi}^{2n-2}+H^{3}_0\mathcal{\psi}^2]
+\frac{2(\gamma-1)(2n-2)^2\Omega_0m}{144(m-5n)\gamma H^{5}_0n^2(1+H^{3}_0n^2)}\mathcal{\psi}^{m-5n}]\coth[\frac{k}{2\mathcal{T}}].
\end{eqnarray}
Figure \textbf{5} represents a parametric plot of $R$ versus $n_s$ for three different values of $n$ and $m=3$.
This plot clearly proves the compatibility of this case with Planck data as both of the perturbed parameters
follow the physical bound for constrained model parameters.

\subsection{Case III:\quad $\Omega=\Omega_0H^2;\quad \xi=\xi_0\rho$}

Under these conditions, the auxiliary function is modified to
\begin{eqnarray}\nonumber
\overline{\chi}(\mathcal{\psi})&=&\frac{-4n\Omega_0H_0}{3(3-n)(2n-2)H^{2}_0n^2
(1+H^{3}_0n^2)}\mathcal{\psi}^{3-n}+\ln[\mathcal{\psi}^{-2}+H^{2}_0n^2
(1+H^{3}_0n^2)\\\label{69}&\times&\mathcal{\psi}^{2n-2}+H^{3}_0\mathcal{\psi}^2]+\frac{(\gamma-1)
(2n-2)^2\Omega_0}{108\gamma H^{3}_0n^2
(1+H^{3}_0n^2)}\mathcal{\psi}^{-3n},
\end{eqnarray}
which further leads us to calculate scalar power spectrum as
\begin{eqnarray}\nonumber
P_{s}&=&\frac{32\mathcal{T}_r\Omega^{\frac{1}{2}}_0H^{-\frac{7}{2}}_0\mathcal{\psi}^{\frac{14-7n}{2}}}
{(2n-2)^2(1+H^{3}_0n^2)^{\frac{5}{2}}n^5}\exp[\frac{8\Omega_0\mathcal{\psi}^{3-n}}{3(2n-2)(3-n)H_0n^2
(1+H^{3}_0n^2)}\\\nonumber&-&2\ln[\mathcal{\psi}^{-2}+H^{2}_0n^2(1+H^{3}_0n^2)\mathcal{\psi}^{2n-2}+H^{3}_0\psi^2]
-\frac{(\gamma-1)(2n-2)^2\Omega_0}{54\gamma H^{3}_0n^2(1+H^{3}_0n^2)}\mathcal{\psi}^{-3n}].\\\label{70}
\end{eqnarray}
\begin{figure}\center
\epsfig{file=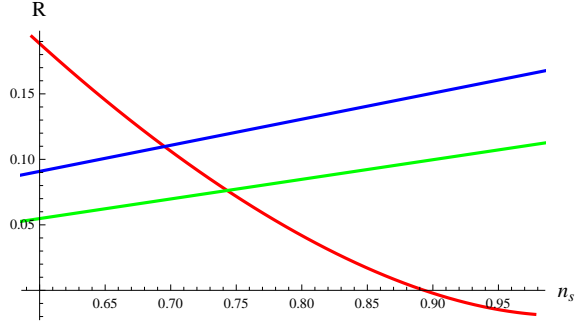,
width=0.55\linewidth}\caption{$R$ versus $n_{s}$: Red for $n=0.5,~H_{0}=3\times10^{-2},~\Omega=0.0022$;
Green for $n=1.5,~H_{0}=9\times10^{-4},~\Omega=0.0002$; Blue for $n=2,~H_{0}=10\times10^{-4},~\Omega=0.0002$.}
\end{figure}
The tensor-scalar spectrum ratio can be calculated as
\begin{eqnarray}\nonumber
R&=&\frac{(2n-2)^2(1+H^{3}_0n^2)^{\frac{5}{2}}n^5H^{\frac{11}{2}}_0}{64\pi^2T_r\Omega^{\frac{1}{2}}_0}
\mathcal{\psi}^{\frac{11n-14}{2}}\exp[\frac{8\Omega_0\mathcal{\psi}^{n-3}}{3(2n-2)(3-n)H_0n^2
(1+H^{3}_0n^2)}\\\nonumber&+&2\ln[\mathcal{\psi}^{-2}+H^{2}_0n^2(1+H^{3}_0n^2)\mathcal{\psi}^{2n-2}+H^{3}_0\mathcal{\psi}^2]
+\frac{(\gamma-1)(2n-2)^2\Omega_0}{54\gamma H^{3}_0n^2(1+H^{3}_0n^2)}\mathcal{\psi}^{-(3n+1)}]\\\label{71}
&\times&\coth[\frac{k}{2\mathcal{T}}],
\end{eqnarray}
and $n_s$ has the following form
\begin{eqnarray}\nonumber
n_{s}-1&=&(\frac{14-7n}{2})\mathcal{\psi}^{-1}+\frac{8\Omega_0(3-n)\psi^{2-n}}{3(2n-2)(3-n)H_0n^2
(1+H^{3}_0n^2)}\\\nonumber&-&2[\frac{-2\psi^{-3}+(2n-2)H^{2}_0n^2(1+H^{3}_0n^2)\mathcal{\psi}^{2n-3}+2H^{3}_0\mathcal{\psi}}
{\mathcal{\psi}^{-2}+H^{2}_0n^2(1+H^{3}_0n^2)\mathcal{\psi}^{2n-2}+H^{3}_0\mathcal{\psi}^2}]
\\\label{72}&+&\frac{(\gamma-1)(2n-2)^2\Omega_0}{18\gamma H^{3}_0n(1+H^{3}_0n^2)}\mathcal{\psi}^{-(3n+1)}.
\end{eqnarray}
The nature of $R-n_s$ trajectory is depicted in Fig.\textbf{6}. For constrained values of the model parameters, the
value of $R$ is always less than $0.11$ for standard value of $n_s=0.968$. Hence the third case of WI model inspired
by tachyon field remains compatible with Planck bound.

\section{Concluding Remarks}

Dissipation is an important phenomenon for the description of entropy mode production. The
inflationary models with viscous effects have ability to generate a rich variety of
power spectra ranging between red and blue. The possibility of a spectrum which runs from
blue to red is particularly interesting, because it is not commonly seen in inflationary
models, which typically predict red spectra. Models of inflation with dissipative effects
and models with interacting fields have much more freedom than a single self-interacting
inflaton in agreement with the observational data.

The inflationary era (a phase of early cosmic evolution) could be gracefully described by
tachyon field, related to unstable D-brane, due to the tachyon condensation near
the maximum of the effective potential. On the other hand, tachyon fields may produced relativistic fluid
or a new type of cosmological dark matter in the cosmos at the late time. Tachyon potentials have two special
characteristics: firstly a maximum of $\mathrm{U}(\psi)$ is obtained, where $\psi\rightarrow0$; secondly
minimum of $\mathrm{U}(\psi)$, which is obtained for $\psi\rightarrow\infty$. If the tachyon field starts
to roll down the potential, then the universe dominated by a new form of matter, will smoothly evolve from
cosmic inflation to an era, which is dominated by a non-relativistic fluid. So, we can explain the phase
of exponential expansion in terms of tachyon field. In the framework of cold tachyon inflation, after slow-roll
phase, tachyon fields evolve towards minimum of $\mathrm{U}(\psi)$ without oscillating about it,
thus, here the reheating mechanism does not applicable. Warm tachyon inflation is a picture, where there are
dissipative effects playing important role during inflation. As a result of this, the inflation evolves in a
thermal radiation bath; therefore the reheating problem of cold tachyon inflation can be solved in the framework
of warm tachyon inflation. It is noted that the cold tachyon inflation era can naturally end with the collision
of the two branes so in this situation, WI is not needed. If the collision of two branes does not arise naturally,
WI is perfectly good scenario that can solve the problem of end of tachyon inflation \cite{HO}.

Motivated by dissipation and tachyon fields, this paper is devoted to discuss warm tachyon inflation with dissipation
and viscous effects originated by tachyon field using a powerful method known as \textit{HJ formalism}. The advantage
of this method is to get rid of too many approximations other than slow-roll used to solve the system of inflationary
model equations. Considering this scenario, we have developed a general criteria to evaluate the solutions of
$\psi$ and $\mathrm{U}(\psi)$ and to modify the slow-roll as well as perturbed parameters for the present
model. Here, the analysis is made in weak dissipative regime. The tachyon inspired inflationary model is
being developed for three different choices of $\Omega$ and $\xi$: $(a)~\Omega=\Omega_0,~\xi=\xi_0$;~
$(b)~\Omega=\Omega_0\mathcal{\psi}^m,~\xi=\xi_0;~(c)~\Omega=\Omega_0H^2,~\xi=\xi_0\rho$, where $m$
is an arbitrary positive constant. The involved model parameters are constrained to get the required results.

The solution of inflaton in terms of number of e-folds is calculated, using this solution, we have formed
the expressions of $\rho_{\mathcal{\psi}},~\rho$ and $\mathrm{U}(\mathcal{\psi})$ as a function of $N$.
The scalar field is slowly rolls down towards minimum value of potential and after
a time inflaton is in equilibrium state as shown in left plot of Fig.\textbf{1}. Right plot of Fig.\textbf{1}
verifies that model is interpolated from high to low dissipative regime.
To observe the nature of these quantities, we have plotted $\rho$ and $\rho_\mathcal{\psi}$ versus $\psi$
in left and right plots of Fig.\textbf{2}. On comparing these two plots, it is noted that imperfect fluid
energy density is much less than inflaton density for specific values of the model parameters. The values
for left plot are constrained to $n=0.5,~H_{0}=1.6\times10^{-2}$ (Dotted curve); $n=1,~H_{0}=5\times10^{-3}$
(Dashed); $n=2,~H_{0}=5\times10^{-4}$ (Thick) while for right plot: $n=0.5,~H_{0}=1.6\times10^{-2}$ (Dotted);
$n=1,~H_{0}=5\times10^{-3}$ (Dashed); $n=2,~H_{0}=5\times10^{-4}$ (Thick). Hence, it can be verified that
the slow-roll condition is true in this scenario. The other involved parameters are fixed to $\gamma=1.5,
~\xi_{0}=7\times10^{-14}$. Further, to check the compatibility of the warm tachyon inflationary model
with observational data, we have plotted $R-n_s$ trajectories for specified values of the model parameters.
Figures \textbf{3,5} and \textbf{6}, plotted for three different choices of $\Omega,\xi$, verify that the model
is in good agreement with Planck bound as $R<0.11$ for $n_s=0.968$ for all the three values of $n$. In Fig.\textbf{4},
the trajectories of $\mathcal{T}_r-H$ are plotted, which proves the existence of WI by satisfying the condition
$\mathcal{T}_r\gg H$.

We have compared the results of our paper with previous literature. It is proved that our model gives more physical
acceptable cases as compared to \cite{29a,JA}. All the three cases (constant and variable coefficients) of tachyon
inspired WI are compatible with Planck data as compared to standard scalar field inflation. The parameters are
more fine-tuned as compared to high dissipative regime \cite{JA}. It is worth mentioning that tachyon inspired WI
with bulk viscous pressure is realistic as its ends gracefully and entered into another cosmic era. In future, we will
discuss this work by implementing first principle of QFT.

\end{document}